\documentclass[runningheads]{llncs}
\usepackage{graphicx}
\usepackage{todonotes}
\usepackage{color,soul}

\usepackage{xcolor}
\usepackage{forest}
\usepackage{multicol}
\usepackage{cite}
\usepackage{amsmath}

\makeatletter
\newcommand{\printfnsymbol}[1]{%
  \textsuperscript{\@fnsymbol{#1}}%
}
\makeatother
\begin{document}
\title{Situation Awareness for Autonomous Vehicles Using Blockchain-based Service Cooperation}  
%
\titlerunning{Situation Awareness for Autonomous Vehicles Using Blockchain Technology}
%
\author{Huong Nguyen\thanks{These authors contribute equally}\orcidID{0000-0001-9067-3396}\and Tri Nguyen\printfnsymbol{1}\orcidID{0000-0001-6483-0829} \and Teemu~Leppänen\orcidID{0000-0002-3513-6106} \and Juha~Partala \orcidID{0000-0001-8181-5604} \and Susanna~Pirttikangas\orcidID{0000-0003-2428-9948}}
%
\authorrunning{Huong Nguyen and Tri Nguyen et al.}
%
\institute{ITEE, University of Oulu, Finland\\
\email{\{firstname.lastname\}@oulu.fi}
}
\maketitle              
\begin{abstract}

Efficient Vehicle-to-Everything enabling cooperation and enhanced decision-making for autonomous vehicles is essential for optimized and safe traffic. Real-time decision-making based on vehicle sensor data, other traffic data, and environmental and contextual data becomes imperative. As a part of such Intelligent Traffic Systems, cooperation between different stakeholders needs to be facilitated rapidly, reliably, and securely. The Internet of Things provides the fabric to connect these stakeholders who share their data, refined information, and provided services with each other. However, these cloud-based systems struggle to meet the real-time requirements for smart traffic due to long distances across networks. Here, edge computing systems bring the data and services into the close proximity of fast-moving vehicles, reducing information delivery latencies and improving privacy as sensitive data is processed locally. To solve the issues of trust and latency in data sharing between these stakeholders, we propose a decentralized framework that enables smart contracts between traffic data producers and consumers based on blockchain. Autonomous vehicles connect to a local edge server, share their data, or use services based on agreements, for which the cooperating edge servers across the system provide a platform. We set up proof-of-concept experiments with Hyperledger Fabric and virtual cars to analyze the system throughput with secure unicast and multicast data transmissions. Our results show that multicast transmissions in such a scenario boost the throughput up to 2.5 times where the data packets of different sizes can be transmitted in less than one second.

\end{abstract}
%
%
%



\section{Introduction}

Automatic driving and connected vehicles have specific requirements for high-quality connectivity \cite{5GAA}. Enhancing the autonomous vehicles' performance can partly be enabled with a clear situational picture constructed from a large amount of environmental context data generated during a journey. These data, e.g., weather, road condition, and data from traffic situations, needs to be delivered with low latency through secure and reliable connections and computing infrastructure. Autonomous navigation and decision-making rely on different sensory systems whose flawless operation is a key factor. Ensuring the technical robustness, as well as the validity of data are also means towards accountability of systems \cite{TrustworthyAI} - developing them are concrete steps on our path towards answering ``Who will be responsible in case of an accident?'', for example. New efficiency and safety designs can increase the resiliency and security of the whole autonomous vehicle systems of systems.


Notably, the real-time data transmission and Intelligent Transportation System (ITS) service access from autonomous vehicles to the backend Internet-of-Thing (IoT) systems is a significant bottleneck due to mobility and physically long distances across networks \cite{Khan2021}. Hence, edge computing as a next-generation IoT platform provides a solution by bringing services, applications, and data into the close proximity of vehicles. In such a decentralized IoT platform, based on local edge servers deployed at a one-hop distance in the (mobile) network infrastructure, data transmission and processing latencies can be significantly reduced and application execution localized. Real-time capabilities for ITS and Vehicle-to-Everything (V2X) systems, in general, are thus improved, for example, related to vehicle sensor data processing and sharing of information between vehicles on the road and access to services by different stakeholders. Moreover, privacy can be enhanced with localized data processing.

As a state-of-the-art technology, blockchain promises the connection among participants in a decentralized environment. To tackle the possible lack of trust among various participants in a decentralized environment, blockchain targets to transparency, immutability, and security and serving as a public distributed database. Trust formation is realised as the confirmed information in blockchain can be witnessed but cannot be modified by anyone~\cite{Queralta2021}.  
Furthermore, the successful formation of a smart contract system is based on blockchain technology. 
With the utilization of the blockchain concept for trust generation in a decentralized system, a smart contract platform distributes execution transactions to system participants for autonomous execution.
Via blockchain and smart contract, the most recent survey\mbox{\cite{mollah_2021_blockchain}} indicates many blockchain's benefits to Internet-of-Vehicles scenarios in ensuring integrity, privacy, fault tolerance, trust, and system connectivity with performance and automation. In regard to the data exchange and management, recent works\mbox{\cite{li2018creditcoin,liu_2018,kang_2019_blockchain}} gain reliability and security in data exchange among vehicles via the deployment of blockchain by Road Side Units (RSUs) or Local Aggregators (LAs). 
However, although these approaches proposed potential architectures and reputation schemes for qualified data evaluation, a question about the network infrastructure is still open. Also, these existing works have not evaluated a Proof-of-Concept (PoC) related to blockchain performance. Finally, these studies have not mentioned the interpretation of information exchange in the decision-making improvement of autonomous vehicles.



With the existing questions, the research aims to leverage a permissioned blockchain in the enhancement of vehicles' decision-making ability within the vehicular network, where data storage, exchange, and service access are based on smart contracts as the agreements between service providers' edge servers in the system.
The proposal is a trusted platform exchanging information among participants\footnote{The terms ``participant'', ``station'', ``node'', and ``peer'' interchangeably use with the same meaning, while the term ``network'' and ``system'' are also interchangeable} through services deployed on edge servers that consume vehicle's, traffic's, weather's, contextual data to produce information related to situation awareness of the road condition, congestion, incidents, and traffic control. In which, distributed services at the edge platform share their information to build a big picture of the traffic and contextual environment that supports autonomous vehicles in their decision-making as well as enables them to adapt the operations according to the situation awareness in real-time. 
%
%
Furthermore, with the lack of a PoC from previous works, this study proposes a scenario for a blockchain-based information exchange system among virtual vehicles in a specific context, taxi companies' cooperation.
We emphasize our contribution to the experiments and simulations that demonstrate the feasibility of our proposed system in the real-world application as a robust and lightweight service delivery for situation awareness among vehicles, with promising low latency in both unicast and multicast transmissions.

The remaining of this paper is organized as: section~\ref{sec:relatedWorks} mentions related works. The proposal and system setup are described at Section~\ref{sec:proposal} and Section~\ref{sec:systemSetup}, respectively. Section~\ref{sec:experiments} is about the experiments and results. Section~\ref{sec:conclusion} concludes the paper through the discussion and future works.


\section{Related Work} \label{sec:relatedWorks}



With the development of ITS, Zhang et al.~\cite{zhang_2011} indicates practical issues for ITS, such as traffic accidents, traffic congestion, and the limitation of land resources. Along with those challenges, a prominent solution considered by~\cite{zhang_2011} is to analyze a large amount of traffic data to form new functions and services in an ITS, for example, the GPS data can support for user behavior prediction. The study therefore envisions the growth of ITS as a data-driven ITS which gathers data as much as possible. Leveraging data for processing in a data-driven ITS, Zhang et al.~\cite{zhang_2011} mentions two major categories, including vision-driven ITS and multisource-driven ITS. In vision-driven ITS, a set of applications includes traffic behavior prediction, traffic object vision related to detection or recognition, and tracking through video sensor. Meanwhile, the multisource-driven ITS consists of sensor data, such as inductive-loop detector, laser radar, and GPS, assist the recognition of environment changes as snow, glare, slippery, and rain. Leveraging traffic information in ITS enables autonomous vehicle systems with efficiency and safety.

Due to the importance of data in both vision-driven and multisource-driven ITS~\cite{zhang_2011}, the evolution of ITS needs a secure, reliable connectivity for data exchange. As the infrastructure for data exchange, an architecture for ITS demands components, including, data generators as vehicles through sensors, edge computing, and data information transmission among stakeholders. To further clarify the gap among the prior literature and how this proposal addresses the problem, hereby, we consider autonomous vehicles in general and the leveraging blockchain in ITS.

Proposed works from the fog and edge layers have been given out to address these latencies \cite{edge-traffic-flow, fog-vehicle-life, pi-edge}. Among edge, fog, and cloud architectures, the key difference is related to deployment servers and data stores across the platform; as computational power increases in layers towards the cloud, so do the latencies across networks. One of the approaches related to our PoC in solving traffic congestion is \cite{edge-traffic-flow}, which took advantages of YOLOv3 and DeepSORT algorithm to detect the traffic flow based on vehicles surveillance.
Recently, Tang et al. \cite{pi-edge} have introduced their $\pi$-edge system as the first complete and super lightweight edge computing system of an autonomous production vehicle, which was successfully evaluated on Nvidia Jetson to support multiple autonomous driving services while only consuming 11W of power.

Security is another issue since autonomous vehicles are vulnerable to attacks in many ways. As regards this, \cite{cyberattack} mentioned all the possible attack categories and cases that can be happened for driverless vehicles. To be more specified, intrusions may occur to automotive control system, such as ECU \cite{Salfer_2014}, driving system components, such as sensors \cite{sensor-att1, sensor-att2}, and during the vehicle-to-everything communications \cite{vanet-att1, infotain-ble-att1}. Besides, outstanding defense approaches in the last decade were also discussed in the paper \cite{bc-review}. According to that, a fused model from CertainLogic and Dempster-Shafer Theory has been introduced by \cite{trustworthiness} to measure the trust value of each participated vehicle in the system considering On-Board Unit components, GPS data, and safety messages.

Related to this work,~\cite{li2018creditcoin,liu_2018} are attracted from the idea of blockchain-based cryptocurrencies for the construction of coins to encourage the information exchange among vehicles. A blockchain-based incentive announcement network is proposed by~\cite{li2018creditcoin} utilizing a reputation point for evaluating information and gaining reputation scores from sharing information. Also, study~\cite{liu_2018} mentions the use of blockchain to build a vehicle sharing system with data coins and energy coins.

Notably,\mbox{\cite{kang_2019_blockchain}} exploited a consortium blockchain for a secure sharing of vehicular data in vehicular edge computing and networks. Edge servers are RSUs that contain and exchange vehicular data in blockchain from vehicles via smart contracts and a reputation scheme for the reliable data source. This work describes the architecture and experiments related to reputation schemes with a dataset but lacks an evaluation of the system's performance. Another point from this study is the utilization of cloud setup and a proof-of-work consensus for a consortium blockchain.
Those usages of blockchain in the vehicle system agree that blockchain maintenance is from RSU and LAs or even official public vehicles to verify and validate blockchain operations, such as consensus tasks.
By exploiting the cooperation among service transport providers based on blockchain, Nguyen et al.\mbox{\cite{nguyen-MaaS}} proposed utilization of blockchain to form a Mobility-as-a-Service with trustworthiness.


Despite many studies for autonomous vehicle systems utilizing blockchain in the field, evaluation and PoC for this study are still questions from our perspectives. Also, instead of forming a setup with connectivity among transportation service providers, for example, taxi companies, the existing studies do not detail to clarify infrastructure for the system; these studies utilize the incentive strategies of blockchain to gain attraction and fairness among participants. Recognize this lack of knowledge; this work proposes a blockchain-based situation awareness for autonomous vehicles, constructs a PoC, and evaluates the throughput.


\section{Our Proposal: Blockchain-based Service Cooperation for Autonomous Vehicles} 
\label{sec:proposal}



In the context of sharing data with a higher standard of safety, along with the need for cooperation among transport vehicles, we propose a secure, reliable collaborating system that takes advantage of the embedded blockchain technology.

\subsection{Blockchain-based Service Cooperation in a Vehicle System}   


\begin{figure}[!]
    \centering
    \includegraphics[width=\textwidth]{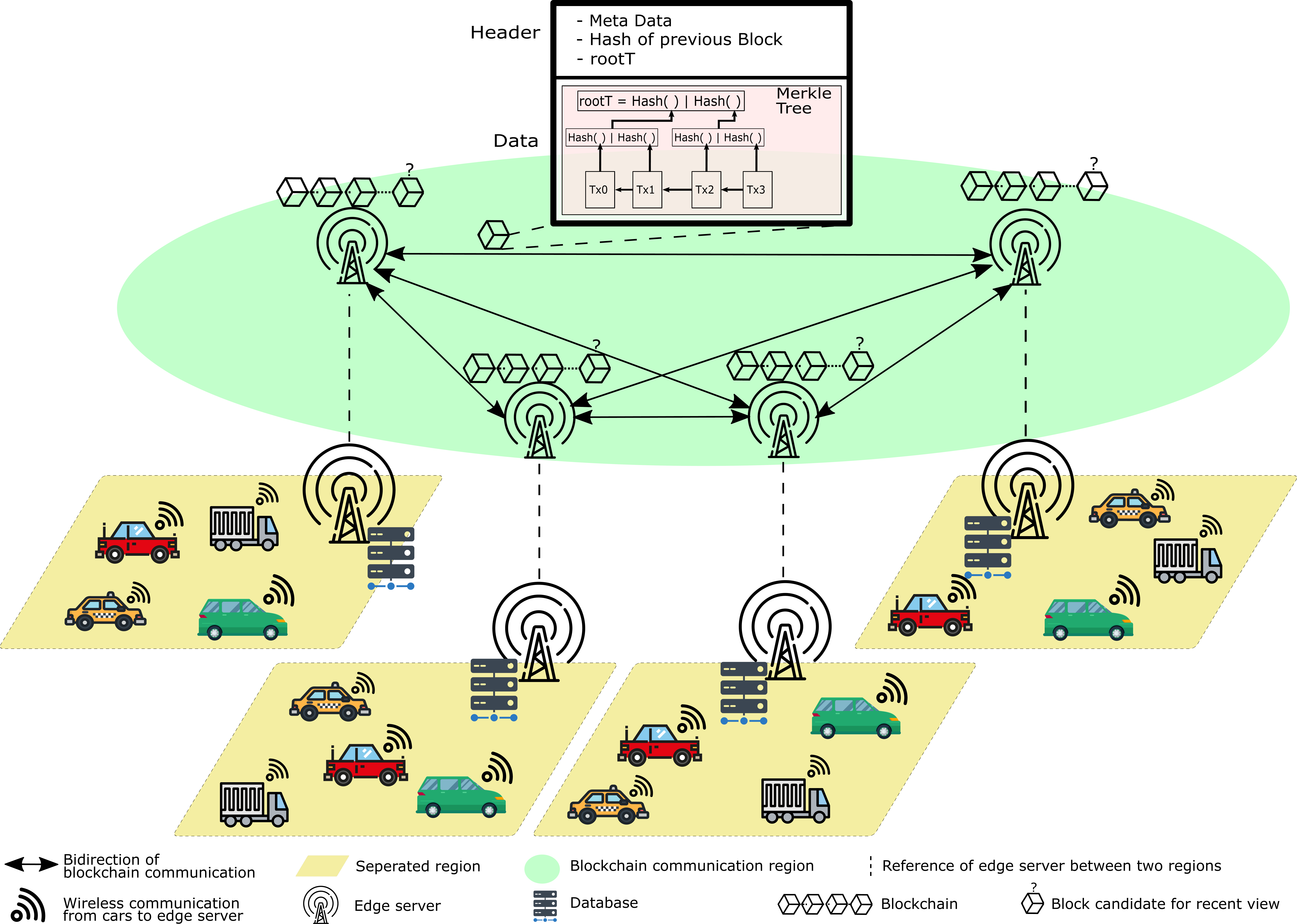}
    \caption{General system design for the proposal}
    \label{fig:generalProposal}
\end{figure}

The proposal is a platform encompassing many types of entities in a transportation scenario to handle the lack of context knowledge from the perspective of autonomous vehicles. 
As a question for the lack of contextual and situational awareness, autonomous vehicles process an amount of data generated by themselves and require much more environmental and traffic information to produce high-performance decisions. Therefore, forming information exchanges among autonomous vehicles is a crucial deliberation; however, this has an obstruction related to trust in the connection among participants. As a prominent solution for trust construction, a distributed deployment of roadside edge servers maintains a unique blockchain by which the data is distributed and stored at every part of the network. 
%

Basically, blockchain is a chain of data blocks, as the green zone in Fig.~{\ref{fig:generalProposal}}, where each block contains two main components: header and data. The header is about metadata, including hashing values of the previous block and Merkle tree\mbox{\cite{merkle1987digital}} to guarantee integrity, while the data part is a list of ordered messages as system's data. Moreover, without intermediaries, a blockchain system requires a consensus mechanism to form agreement and validation on blockchain proposals; for example, at the green zone in Fig.{~\ref{fig:generalProposal}}, four edge servers communicate and validate the block candidate.

%
Thus, the system provides trust among the participants to access the vehicle data via blockchain's secure, transparent, and immutable sharing mechanism. Moreover, although the system's failure can be from data transmission or even crashed participants, blockchain technology enables the fault tolerance for the system based on a consensus mechanism. 
With the distributed data, an edge server at a specific location shares to vehicles within its coverage. Those servers handle computational tasks before emitting recommendations to vehicles in the local environment. As local edge servers facilitate short-range communications, less latency is experienced, and less transmission power is consumed when data is processed locally before transmitting or storing it with distributed blockchain technology. In addition, processing the data on edge maintains privacy for the system.   
Consequently, the collaborative work between blockchain and edge can lead to a promising, fast, and effective data storage and sharing platform in real-time.

With this perspective, the general architecture proposal encompasses two main layers, including blockchain communication and communication between cars and edge servers as green and yellow zones in Fig.~{\ref{fig:generalProposal}}, respectively.
Referred to Fig.~{\ref{fig:generalProposal}}, our proposed system provides a blockchain-based infrastructure as a channel for collaborative communication among service providers. In detail, the system separates into an array of regions as yellow sides in Fig.~{\ref{fig:generalProposal}}. An edge server manages each region as a service provider, which consists of three main tasks: (1) collecting vehicles' information, (2) interpreting from vehicles' information, and (3) maintaining a unique blockchain with other service providers. With the consideration, vehicles at a specific region share captured data related to the contextual environment, for example, road conditions or traffic information, to the edge server of the region via pre-defined smart contracts. 
Although the contextual environment is vision and multisource data, we primarily  focus on vision information through this work. For example, a vehicle can capture images of the surrounding environment and send them to the edge nearby with its identity and GPS as location information.
After gathering vehicles' information, the edge server analyses to understand the region's situation and broadcasts the knowledge to other edge servers. 
Consequently, the information of each yellow region is broadcast to entire edge servers and waits for wrapping and validating into a block as a new confirmed state. Once accepting a new state, edge servers send contextual information and knowledge of the new state to vehicles in their region to enhance the decision-making ability of the vehicles.
In other words, the edge servers as a set of service providers maintain a unique blockchain storing vehicles' sensor information and interpreting knowledge of any participants.

An interest for a blockchain-based vehicle system is a wide range of services. These can be deployed with the smart contract concept as a platform that supplies standards and infrastructure through network participation. For example, with a blockchain-based smart contract for insurance, the proposal can provide and construct an insurance system. Once existing accidents, the system automatically collects evidence and sends specs to the insurance company. These shreds of evidence can include the log of the car's speed, the latest image before the incident, the latest car inspection time, or the history of owners.

A blockchain-based system could be a promising solution to the problem of resource distribution. For example, when numerous cars in a zone send their acquired data to the same server, the workload turns heavy, and it becomes bottlenecked while the others are idle. The possible answer for this issue is based on the rank of the received requests, by which the peer forwards tasks with low priority to the neighbors for assistance in offloading. For example, following~\cite{Dimitrakopoulos_2010}, the definition of high-level data is the knowledge related to emergencies, congestion, drivers' behaviors, vehicles' condition, roads' condition, and neighboring vehicles. Meanwhile, the low-level data is about accurate position, drivers information, and vehicles information.


Whenever data is stored in the blockchain, its correctness needs to be ensured by the validating nodes. Due to such public verifiability requirements, data privacy poses a significant problem for any blockchain application that stores private data. Regarding the storage of vehicular information, one needs to be very careful not to violate the privacy of the drivers and passengers. Therefore, anonymization techniques need to be applied to remove all personally identifiable information.
In our system, all information is pseudonymized. Each vehicle is identified by its unique public key and no other identification information is stored. Furthermore, requests from the vehicles to the edge servers need to be encrypted and mutually authenticated in order to prevent eavesdropping and impersonation.
For smart contract applications that need
to store personal data, zero-knowledge arguments can be used to mitigate privacy issues by showing the correctness of statements
without disclosing any other information. There are non-interactive zero-knowledge-based schemes, such as Hawk~\cite{Kosba_2016} and Zexe~\cite{Bowe_2020}, that can be used to protect blockchain-based smart contracts and enable them to encrypt data. For a survey and comparison of such schemes, see for example~\cite{Partala_2020}.
Such methods need to be applied together with anonymization techniques to remove any personally identifiable information before data is stored in the system.

Instead of considering the infrastructure providers, the proposal system is a collaborative platform, attracting service providers through their fascination. In collaboration with an array of edge servers, the infrastructure for the system is a question. Thus, the system can cooperate among different organizations, such as transport service providers, to maintain the blockchain. The proposed system is a collaborative channel that permits organizations to provide their infrastructure and define the agreement as supporting services. Also, instead of forming a cluster of vehicles through connecting smart vehicles, the edge server as the central point provides the channel gathering smart vehicles in the locality with robust computation, reliability, and security.

\subsection{Proof-of-Concept in Situation Awareness for a Blockchain-based Taxi Service}

    \begin{figure}[!]
    	\centering
    	\includegraphics[width=0.89\textwidth,height=7.59cm]{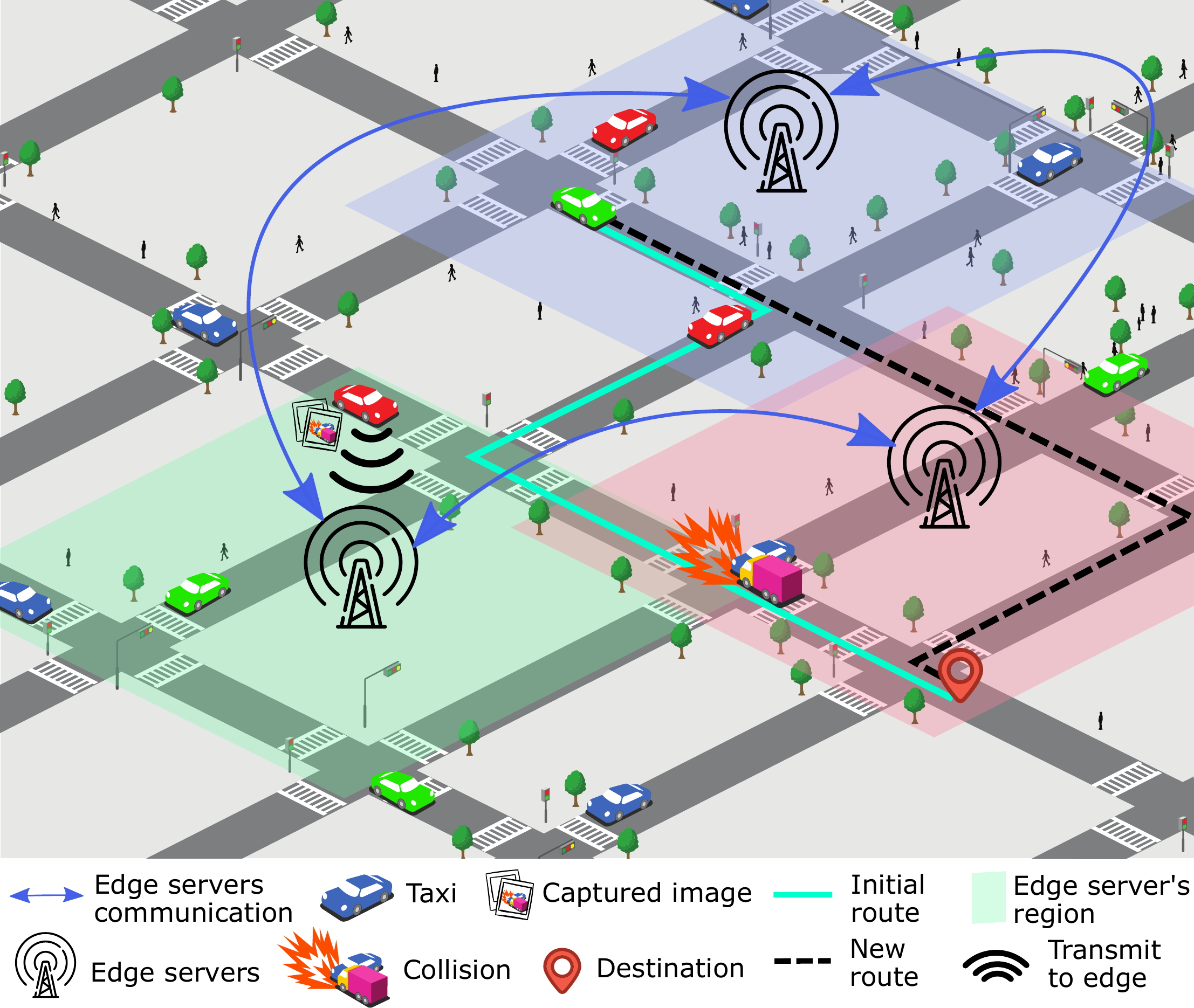}

    	\caption{Proof-of-Concept for blockchain-based taxi service: adaptive guidance}
    	\label{fig:poc-usecase}
    \end{figure}
For the PoC, we aim to create a friendly sharing environment for a specific group of targets: the taxi. In particular, our platform encompasses multiple taxi companies as participants to join and share their data. To be more specified, two or more taxis will share their locations and image data to the nearest edge server, which is known as the base station, without concern about which taxi company the server belongs. 

According to that, data after being stored in one server will be replicated to others within the network, which belongs to other companies. This broadcasting mechanism enables every car throughout the cooperated network to have the access to all data almost simultaneously and accurately. Cars are also able to be constantly aware of critical factors affecting the transportation process, such as road conditions based on other cars' status. For example, we can know about the current location's situation from the image data, understand the surrounding context, and process the information to detect early congestion or unexpected incidents. 
Consequently, a pre-defined route to a destination in a car will be automatically changed to another if there is a congestion or any incidents detected. The voice navigation is updated accordingly to guide the drivers to safely switch the route to reach the same target. Shortly, adaptive routing will be an indispensable feature for cars, compatible, easy to integrate with multiple in-car applications. Hence, it becomes beneficial for the drivers to avoid unexpected events and have a smoother road trip towards the ITS.

We present an illustration for the use case of adaptive guidance in Fig.~{\ref{fig:poc-usecase}}. Three edge servers placed in the same-color zones respond to three taxi companies: red, green, and blue. As can be seen, a taxi from the red company has witnessed an accident on its way of travel and later sent the captured image along with the GPS data \textit{(longitude, latitude)} to the base station in its zone \textit{(the green server)}. Meanwhile, another car from the green company, being in the blue zone and trying to reach the destination, will instantly receive an updated message from the blue's server. This car will then change its pre-defined route \textit{(solid turquoise line)} to a new one \textit{(dashed black line)} to avoid accidents and congestion. This process can be done due to the broadcasting from green's server to all the participants, including the blue one, shortly after receiving the captured image from the red car. As a result, all vehicles working within the blue zone will be notified about the collision. Furthermore, the integrated routing application in each car will automatically adjust its route in response to the impact of the occurrence.

\section{Blockchain-based Service Cooperation Setup}  \label{sec:systemSetup}




\begin{figure*}[!h]
    \centering
    \includegraphics[width=\textwidth,height=5cm]{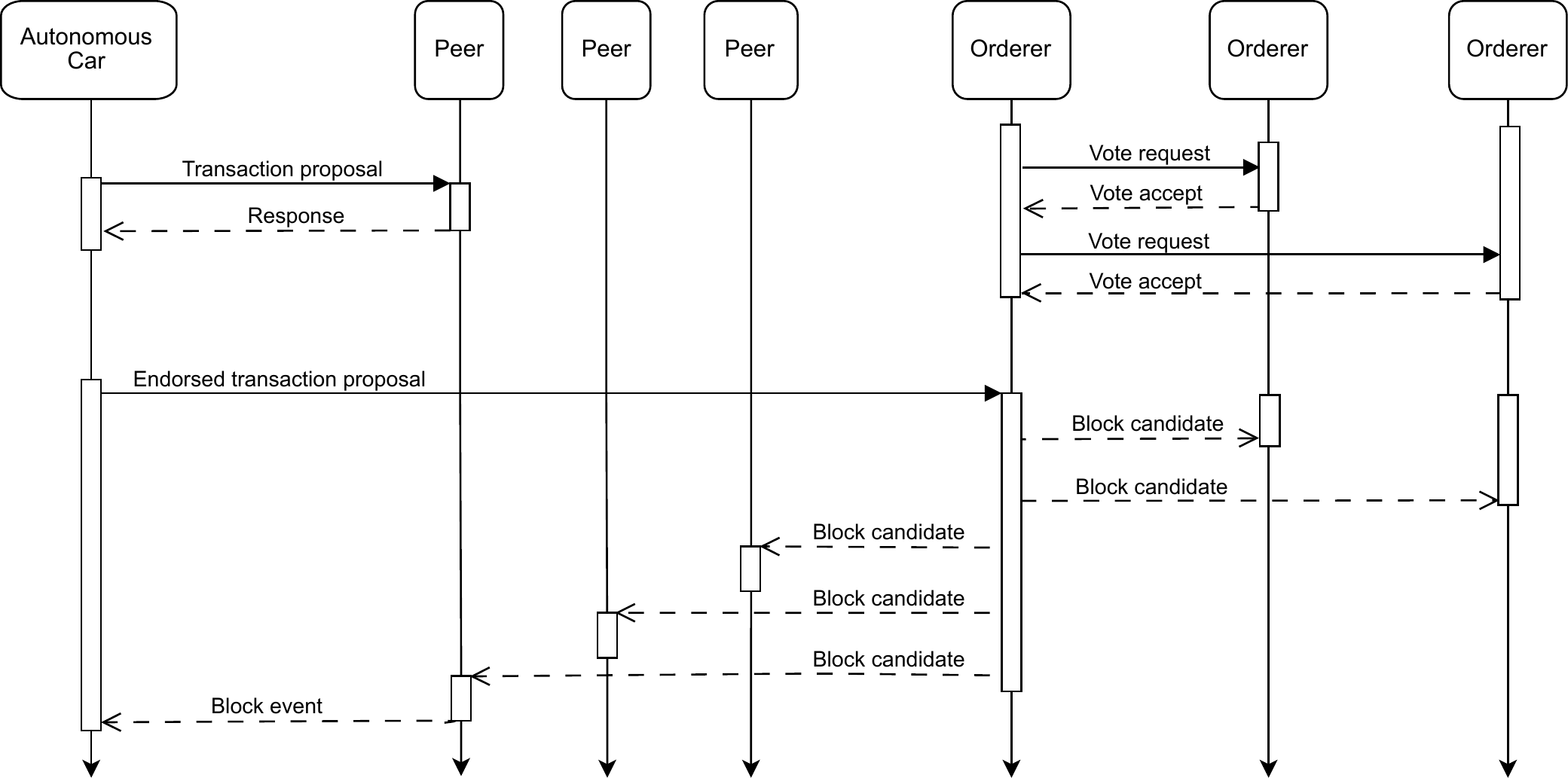}
    \caption{Diagram for leveraging blockchain into an autonomous car system} \label{fig:sequenceDiagramFlow}
\end{figure*}

    The simulation for the blockchain is based on a permissioned blockchain platform called Hyperledger Fabric\footnote{www.hyperledger.org/projects/fabric}. As two primary services for Hyperledger Fabric~\cite{androulaki2018hyperledger}, \textit{Peers} are to maintain a unique ledger as blockchain data and smart contract environments. 
    In detail, the peers have two roles, including endorser and committer. The endorser task is to simulate transactions and prevent non-deterministic transactions while committer peers are appending validated transactions. Notice that a peer can act in both roles. With this consideration, the peer from our setup carries both two roles.
    In contrast, \textit{Orderers} are responsible for arranging endorsed transaction proposals to form a block candidate as the consensus operation. 
    Also, as a permissioned blockchain platform, the system leverages a certificate authority to provide certificates to participants as identity; hence, the consensus for the permissioned blockchain is based on traditional aspects, such as Paxos or Raft.
    Additionally, Fabric proposes a procedure called execute-order-validate that describes a system that first receives requests from clients for handling operators at peers before submitting to orderers to arrange and form a data block. This block is then broadcast to all system's peers for validating at the validation phase before updating the blockchain.    
    As the PoC for the blockchain-based vehicles proposal, a simulation is set up as Fig.~\ref{fig:sequenceDiagramFlow} with several peers and orderers handling a permissioned blockchain system. 
    
    Our system's setup needs to encompass at least three peers and three orderers to simulate a blockchain system that supposes to handle a single failure from a system; meanwhile, the number of autonomous cars can vary. Therefore, if one of the peers or orderers crashes with any issues, the other can alternate the tasks. Regarding the consensus mechanism in our simulation, Raft~\cite{ongaro_2014} is deployed on three Orderers to arrange for the chronological order of requests. As a brief description for the setup as Fig.~\ref{fig:sequenceDiagramFlow}, several system's peers, in the beginning, gather information and requests generated by vehicles for executing. After that, these peers return the vehicle's execution results, forming endorsed transaction proposals and sending them to orderers. Finally, the orderers collect those requests and arrange them into a block candidate for the specific consensus round before broadcasting to all system's peers.
    %
    
    \begin{figure}[t!]
	\centering
	\includegraphics[width=6cm,height=5cm]{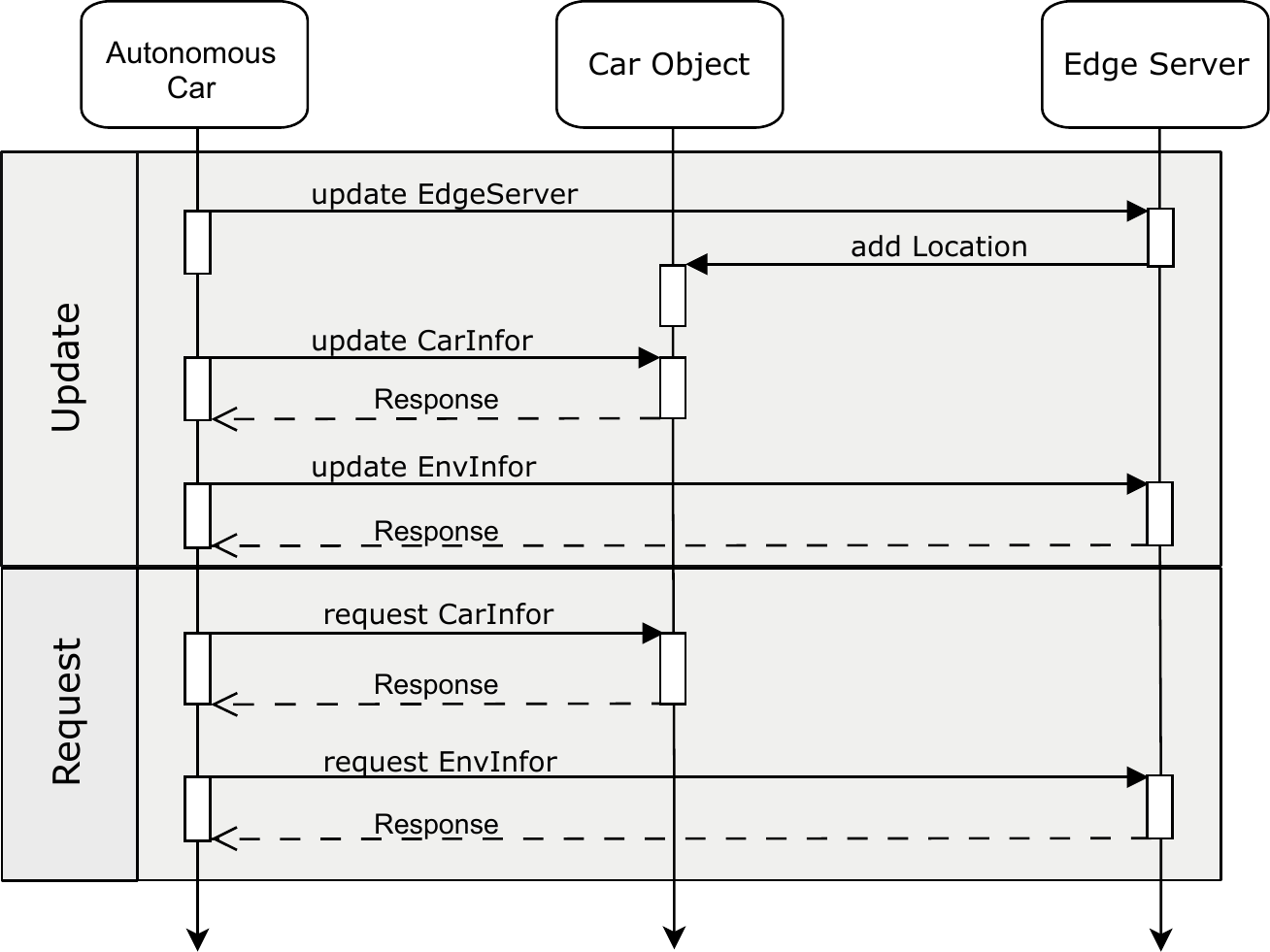}
	\caption{The sequence diagram of the smart contract in the proposal}
	\label{fig:smartContract}
    \end{figure}
    
    With the PoC scenario, a set of smart contracts are deployed as services provided by the system. Following Fig.~\ref{fig:smartContract}, the system receives requests that update information from the autonomous vehicles. 
    For example, the smart contract's update information includes the sequence of owners, inspection history, GPS, currently connected edge server, and even insurance information. Also, the car captures data related to the surrounding environment for updating the local state to the edge nearby, such as images or information about the congestion, traffics, road conditions, and weather.
    On the other hand, the request of information consists of two main procedures: revealing the car information and requiring for environmental information. Therefore, the system swiftly reacts to cars' requests; for example, once an accident exists between two vehicles, the accident information can be transmitted to the system that autonomously contacts the insurance of the violating vehicle. Further, if the edge server receives any urgent situations, such as congestions or accidents, the system directly emits notifications to cars nearby.
    
    
    
\section{Experiments and Results}   \label{sec:experiments}
The main idea of experiments is to evaluate the throughput of our proposal with a simulation to demonstrate feasibility and practicability. Due to the demand for communications among servers in a decentralized environment, throughput is an important performance indicator. Further, throughput has not been considered in the related work.


    \subsection{Resource Configuration}
    
    The setup for the blockchain network is based on six virtual machines (three for Peers and three for Orderers)  as Fig.~\ref{fig:sequenceDiagramFlow}. Those virtual machines are deployed on a physical machine with Intel(R) Core(TM) i7-8700 CPU @ 3.20GHz, 3.19GHz, 64-bit OS, x64-based processor, and 32GB memory through a virtual machine program called Virtual Box Version 5.2.22 r126460 (Qt5.6.2). 
    Meanwhile, the experiments consider three virtual vehicles, for each is set up on a virtual machine with 4 GB memory and 4 cores in processors, as described in Table~\ref{tab:harwareNetwork}. 
    Virtual machines deploy some Hyperledger Fabric's containers, such as fabric-orderer, fabric-peer, fabric-ca, and fabric-tools, through a docker version 18.09.7. In our particular setting, the edge servers deploy a peer, whereas the virtual vehicles utilize Fabric's tools to interact with peers. The orderer service can be deployed in the same machine with a peer or separated, but we decide the separation. The sizes of multiple microservices are presented in Table{~\ref{tab:harwareNetwork}}, especially the fabric-ca' size being 157.9 MB. In particular, peer machines deploy fabric-peer via docker container while fabric-orderer is set up on orderer machines.
    Moreover, one of the orderer machines maintains a certificate authority for the system via fabric-ca container. Similarly, a peer machine uses fabric-ca to distribute certificates to other peers. Besides, virtual cars utilize fabric-tools to interact with peers and orderers.

    
    \begin{table}[]
    \centering
    \caption{Detailed resources and containers for system's components} \label{tab:harwareNetwork}
    \begin{tabular}{|l|l|l|l|}
    \hline
                & Resources                & Container Info               & Arch     \\ \hline
    Orderer     & Process: 2, Memory: 2GB  & fabric-orderer:2.2.1 - 38.42MB & x86\_64   \\ \hline
    Peer        & Process: 4, Memory: 6GB  & fabric-peer:2.2.1 - 49.41MB    & x86\_64   \\ \hline
    Virtual Car & Process: 4, Memory: 4GB  & fabric-tools:2.2.3 - 513.30MB  & x86\_64   \\ \hline
    \end{tabular}
    \end{table}

\subsection{Simulation}


    As an experiment for throughput of the simulated system, virtual vehicles generate numerous requests to edge servers with results in Table~\ref{tab:edge-scalability}. Notice that since we utilize three Peers and three Orderers as the edge servers, the experiments separate into \textit{single} and \textit{multiple} standing for a request sending to one Peer and a request sending to three Peers as unicast and multicast, respectively. Also, the experiment has diverse requests capacity, for example, 16 KiB, 32 KiB, 64 KiB, 100 KiB in each request. By different capacities, each virtual vehicle sends a sequence of 1000 requests to a Peer; hence, with single communication at Table~\ref{tab:edge-scalability}, the total requests that the system has to handle are 3000, whereas this figure in multiple communication is 9000. 
    The experiment indicates the number of transactions that the system can process in a second. Especially, \textit{multiple} experiment's results are double as \textit{single}'s, and even gains more 2.5 times than \textit{single} result with the request capacity of 100 KiB.
    

    Another consideration from the throughput is the notification to update the change from the edge server to vehicles. In this case, we deploy an experiment in which the throughput is calculated by the total time from the update request called by a virtual car to another via the confirmation of edge servers. Like the previous setup, this experiment is based on a diversity of capacity, including 16 KiB, 32 KiB, 64 KiB, and 100 KiB for capturing the throughput. In \textit{single}, a virtual car calls 1000 update requests to the system that notifies those changes to another virtual car, whereas, with \textit{multiple}, the system receives multicast transmissions from a virtual car before notifying another. The results in Table~\ref{tab:contextAwareness} indicate the higher capacity leading to the lower amount of transactions that the virtual car can receive in both unicast and multicast. Similar to the aforementioned experiment, the results of \textit{multiple} double the \textit{single}; however, at the requests with 100 KiB, the gap between \textit{single} and \textit{multiple} is smaller than the others.
    
    An interesting observation from two Table~\ref{tab:edge-scalability} and Table~\ref{tab:contextAwareness} indicates the high amount of processed KiB per second with the high capacity carried by transactions. In detail, the increase of capacity that transactions carry is the higher amount of KiB/s; meanwhile, that leads to the lower number of transactions are processed by the system. For example, in \textit{single} part of Table~\ref{tab:edge-scalability}, the number of transactions 16 KiB is averagely processed about 1.973 in a second, which regards to a mean of 31.568 KiB/s stored by the system. On the other hand, with transactions higher capacity as 32 KiB, 64 KiB, and 100 KiB, the average amount of data that the system stores is 59.328, 92.928, and 106.126, respectively. At the same time, the number of transactions processed by the system reduces according to the higher capacity carried by transactions; particularly, those are 1.973, 1.854, and 1.061 regarding the transactions' capacities 16 KiB, 32 KiB, 64 KiB, and 100 KiB. This consideration is similar to others, including \textit{single} part of Table~\ref{tab:contextAwareness}, \textit{multiple} part of Table~\ref{tab:edge-scalability} and Table~\ref{tab:contextAwareness}. 
    
    \begin{table}[h!]
        \caption{The throughput of our proposed system with various capacities of the requests in two types of communication: singe and multiple as unicast and multicast transmission}
        \label{tab:edge-scalability}
        \centering
            \begin{tabular}{|l|cccl|cccl|}
            \hline
                  & \multicolumn{4}{c|}{Single} & \multicolumn{4}{c|}{Multiple} \\ \hline
            Capacity (KiB) &
              \multicolumn{1}{c|}{16} & 
              \multicolumn{1}{c|}{32} &
              \multicolumn{1}{c|}{64} &
              100 &
              \multicolumn{1}{c|}{16} &
              \multicolumn{1}{c|}{32} &
              \multicolumn{1}{c|}{64} &
              100 \\ \hline
            \#transactions/s &
              \multicolumn{1}{c|}{1.973} &
              \multicolumn{1}{c|}{1.854} &
              \multicolumn{1}{c|}{1.452} &
               1.061 &
              \multicolumn{1}{c|}{4.637} &
              \multicolumn{1}{c|}{3.665} &
              \multicolumn{1}{c|}{2.785} &
              2.559
               \\ \hline
            \#KiB/s &
             \multicolumn{1}{c|}{31.568} &
              \multicolumn{1}{c|}{59.328} &
              \multicolumn{1}{c|}{92.928} &
               106.126 &
              \multicolumn{1}{c|}{74.192} &
              \multicolumn{1}{c|}{117.280} &
              \multicolumn{1}{c|}{178.240} &
              255.907 \\ \hline
            
            \end{tabular}
    \end{table}





    \begin{table}[h!]
        \caption{Situation awareness's throughput with various capacities of requests in two types of communication: singe and multiple as unicast and multicast transmission}
        \label{tab:contextAwareness}
        \centering
            \begin{tabular}{|l|cccl|cccl|}
                \hline
                \multicolumn{1}{|c|}{} & \multicolumn{4}{c|}{Single}                                                                                       & \multicolumn{4}{c|}{Multiple}                                                                                     \\ \hline
                Capacity (KiB)          & \multicolumn{1}{c|}{16}     & \multicolumn{1}{c|}{32}    & \multicolumn{1}{c|}{64}    & 100                       & \multicolumn{1}{c|}{16}     & \multicolumn{1}{c|}{32}    & \multicolumn{1}{c|}{64}    & 100                       \\ \hline
                \#transactions/s        & \multicolumn{1}{c|}{1.506} & \multicolumn{1}{c|}{1.328} & \multicolumn{1}{c|}{1.107} & 0.953                    & \multicolumn{1}{c|}{3.271} & \multicolumn{1}{c|}{2.807} & \multicolumn{1}{c|}{2.383} & 1.538                     \\ \hline
                
                \#KiB/s &
                \multicolumn{1}{c|}{24.096} &
                \multicolumn{1}{c|}{42.496} &
                \multicolumn{1}{c|}{70.848} &
                95.336 &
                \multicolumn{1}{c|}{52.336} &
                \multicolumn{1}{c|}{89.824} &
                \multicolumn{1}{c|}{152.512} &
                153.851 \\ \hline
              
                \#s/transaction        & \multicolumn{1}{c|}{0.66}   & \multicolumn{1}{c|}{0.75}  & \multicolumn{1}{c|}{0.94}  & \multicolumn{1}{c|}{1.05} & \multicolumn{1}{c|}{0.31}   & \multicolumn{1}{c|}{0.36}  & \multicolumn{1}{c|}{0.42}  & \multicolumn{1}{c|}{0.65} \\ \hline
            \end{tabular}
    \end{table}
    





\section{Discussion and Future Works}    \label{sec:conclusion}


The paper proposes a blockchain-based vehicle system that supports situation awareness for vehicle decision-making. Interestingly, the system provides a smart contract platform; hence, an array of services for a vehicle system can be deployed in our proposal. By leveraging blockchain technology, the system achieves trust in the collaboration among parties and fault tolerance in a vehicle system. Also, the concept of edge computing is to manage and handle data as close as the data capture, which contributes towards a real-time ITS/V2X system. From this perspective, the proposal is a blockchain-based system encompassing a set of edge servers furnishing real-time decision-making based on situation awareness of vehicles' locality and transparent history.

The experiments evaluate the throughput for the proposal with the unicast and multicast transmission. The results of the experiments indicate the auspicious proposal with the high number of processed transactions per second, especially the multicast transmission obtaining double in comparison with the unicast transmission. 
Notably, although the number of transactions that the system can handle is less in the transactions' large capacity than transactions having lower capacity, the total amount data in a second (KiB/s) from processed transactions with high capacity is higher. In addition, due to the decentralized environment that broadcasts most information, the numerous participants in both vehicles and edge servers impact the system performance. 
Especially in the scalability aspect of a decentralized system, like the proposal, a raised question is about ensuring the system's performance with different numbers of participants who synchronize client requests to gain the agreement of validation and confirmation. Therefore, this perspective is a dilemma for decentralized systems. Interestingly, despite the decentralization, blockchain-based service cooperation leverages a permissioned blockchain that inquires participants' permission and separated organizations before becoming a part of the system, which leads to a restriction for exploding the significant number of participants. Further, blockchain-based service cooperation indicates specific services limited to specific organizations can supply.

Since the proposal experiments' components are virtual machines, some gaps connect to realistic deployments. Firstly, virtual cars are stable without movement, especially interchange between two regions. Another gap from the proposal is the limitation of resources from edge servers and virtual cars. Further, although the blockchain setup foundation is on different machines, the setup configuration is a centralized station without detailed consideration of distance connection among edge servers.
Therefore, the simulation indicates a reflection of an actual model instead of considering the accurate result for deployment. Besides, a fully autonomous vehicle needs to be liable for any at-fault collisions under existing automobile products liability laws, but its ethical values are still debated.

For a practical application of our system, privacy issues need to be thoroughly investigated. Currently, all vehicles are identified by their pseudonyms and no other identification information is stored. However, it is a significant threat that individuals and their movement patterns could be inferred from the data stored on the blockchain. Therefore, any personally identifiable information and other information that could lead to the identification of individuals needs to be purged from the data before it is stored in the system. Zero-knowledge argument schemes can be applied to encrypt data stored in the smart contracts while simultaneously providing assurances of its correctness.
Differential privacy can also be used to prevent the identification of individuals and to anonymize data that cannot be encrypted.
However, in this paper, our focus is on the performance of our proposal, and, therefore, such considerations are left for future work.

\section*{Acknowledgment}
This research is done in a strategic research project {\it TrustedMaaS} under focus institute Infotech Oulu, University of Oulu, and ECSEL JU FRACTAL (grant 877056). A personal grant by the Nokia foundation for Mr. Tri Nguyen. The researchers operate under Academy of Finland, 6G Flagship (grant 318927).

\bibliographystyle{splncs04}
\bibliography{main}

\end{document}